**Title:** Learning Bundled Care Opportunities from Electronic Medical Records

**Authors:** You Chen[1], Abel N. Kho[2], David Liebovitz[3], Catherine Ivory[4], Sarah Osmundson[5], Jiang Bian[6], and Bradley A. Malin[1,7, 8]

**Author Affiliations:**

[1]Dept. of Biomedical Informatics, School of Medicine, Vanderbilt University, Nashville, TN

[2]Institute for Public Health and Medicine, Northwestern University, Chicago, IL

[3]School of Medicine, University of Chicago, Chicago, IL

[4]School of Nursing, Vanderbilt University, Nashville, TN

[5]Dept. of Obstetrics and Gynecology, School of Medicine, Vanderbilt University, Nashville, TN

[6]Dept. of Health Outcomes and Policy, University of Florida, Gainesville, FL

[7]Dept. of Biostatistics, School of Medicine, Vanderbilt University, Nashville, TN

[8]Dept. of Electrical Engineering & Computer Science, School of Engineering, Vanderbilt University, Nashville, TN

**To Whom Correspondence Should Be Addressed:**

You Chen, Ph.D.
Assistant Professor
Department of Biomedical Informatics
Vanderbilt University
Nashville, TN 37203 USA
Email: you.chen@vanderbilt.edu
Phone: +1 615 343 1939
Fax: +1 615 322 0502

**Highlights**

(1) A data-driven approach to learn bundled care opportunities from electronic medical records.

(2) A strategy to infer association network of phenotypic and workflow patterns.

(3) An evaluation of bundled care opportunities with administrative and clinical experts.

**Graphic Abstract**

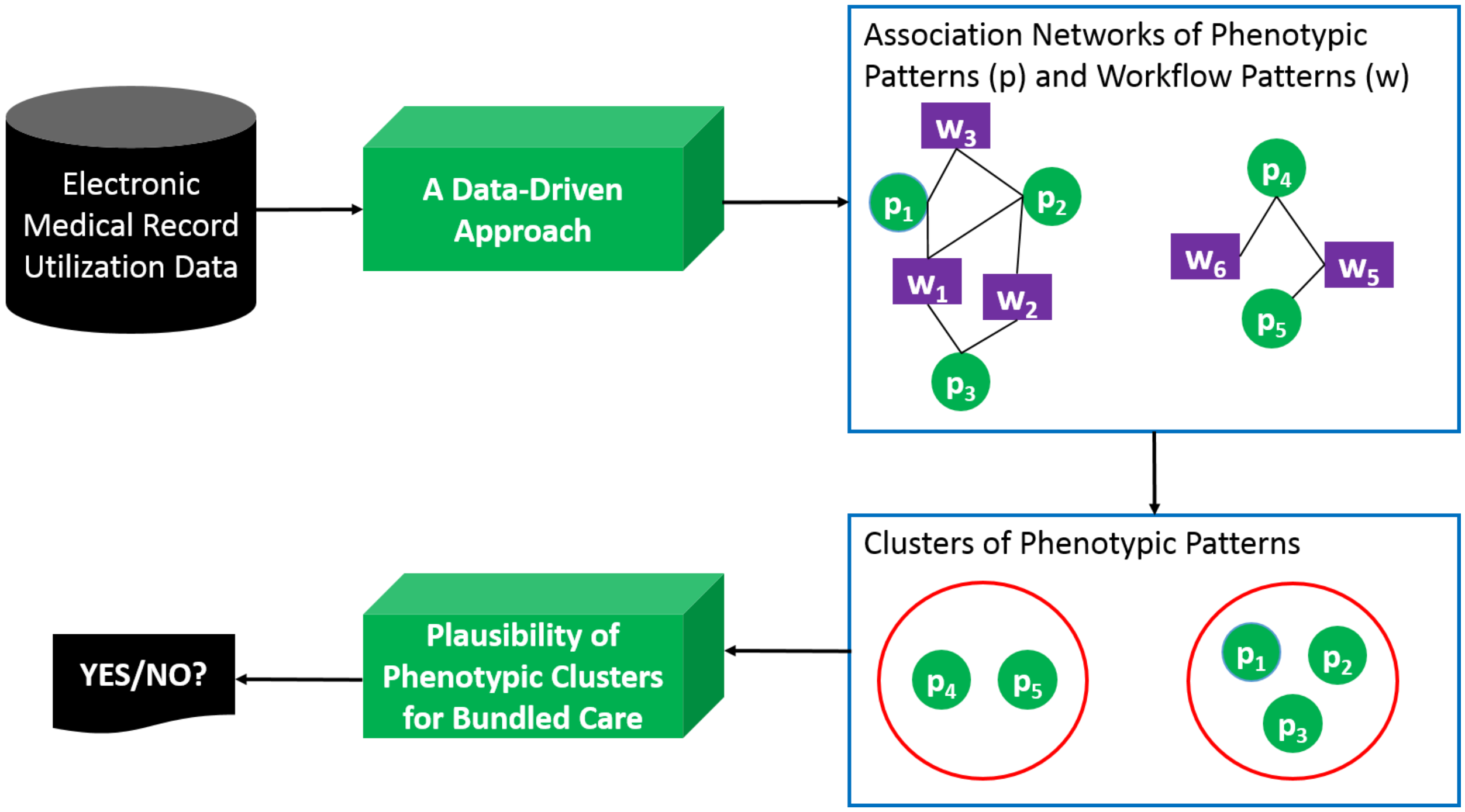

## ABSTRACT

**Objective:** The traditional fee-for-service approach to healthcare leads to the management of a patient's conditions in an independent manner, inducing various negative consequences. It is recognized that a bundled care approach to healthcare - one that manages a collection of health conditions together - may enable greater efficacy and cost savings. However, it is not always evident which sets of conditions should be managed in a bundled manner. Thus, we investigated how a data-driven approach could be applied to automatically learn potential bundles and evaluate their plausibility.

**Methods:** To accomplish this research, we designed a data-driven framework to infer clusters of health conditions, which we referred to as phenotypic patterns, via their shared clinical workflows, which we refer to as coordinating care patterns, from the data inherent in electronic medical records (EMRs). We applied the framework with approximately 16,500 inpatient stays from a large medical center. The plausibility of the inferred health condition clusters for bundled care was assessed through a survey (whose responses were analyzed via an analysis of variance (ANOVA) under a 95% confidence interval) of a panel of five experts. Furthermore, the face validity of the inferred health condition clusters was confirmed by evidence in the published literature.

**Results:** The framework inferred four condition clusters: 1) fetal abnormalities, 2) late pregnancies, 3) prostate problems, and 4) chronic diseases (with congestive heart failure featuring prominently). Each cluster had evidence in the literature and was deemed to be plausible for bundled care via ANOVA on the survey responses under a 95% confidence interval.

**Conclusions:** The findings suggest that data from EMRs can provide a basis for discovering new directions in bundled care. Still, translating such findings into actual care management will require further refinement, implementation, and evaluation.

**Background and Significance**

Under a fee-for-service healthcare model, each of a patient's conditions is managed relatively independently [1-2]. This approach to care can lead to several problems, including delays in (or failure to deliver) service, treatment redundancies, and increased cost. In turn, these problems can lead to declines in quality, patient satisfaction, and cost effectiveness. It is anticipated that a shift from fee-for-service to pay-for-value has the potential to resolve (or at least reduce the severity of) many of these problems [3-5]. To realize this alternative vision, healthcare organizations (HCOs) are migrating towards a bundled care model, which is a middle ground between fee-for-service and capitation reimbursement, and aims to account for the interplay between various health conditions, rather than focus on each in isolation [6-7].

There are numerous challenges in realizing bundled care. Two of the more pressing are: 1) determining which health conditions would be appropriate for such care models and 2) minimize the cost of refining current healthcare systems to support bundled care. While HCOs already manage the complicated health needs of their patients (e.g., considering a set of health conditions together), such routines often arise in an *ad hoc* fashion and are not formalized and validated. As such, there is an opportunity to design a data-driven approach to learn collections of health conditions, which are managed together (e.g., a set of health conditions share similar workflows) and, thus, might be ripe for bundling. The data-driven approach may further be beneficial because, if models are based on current healthcare systems, HCOs could minimize the implementation costs of newly established, or the formalization of existing, management routines.

There is growing evidence that data derived from electronic medical records (EMRs) can be leveraged to discover associations between health problems [8-14], infer clinical phenomena (e.g., phenotypic patterns [15-18), and model workflows (e.g., hospital care patterns [19-23). More recently, it has been shown that the relationship between health problems and workflows can be learned for specific phenomena, such as congestive heart failure [24]. In this paper, we build on such observations and introduce an automated learning framework to discover more general collections of health conditions that share similar workflows according to EMR system utilization. We hypothesize that such collections of health conditions could be bundled and managed together based on their shared workflows.

We accomplish this goal by applying a generative topic modeling strategy to infer phenotypic patterns from the data inherent in EMRs and workflow patterns from the utilization of EMRs by employees of a healthcare organization (HCO). We apply a community detection algorithm to infer clusters of phenotypic patterns that share workflow patterns. We evaluate this framework with four months of inpatient data (over 16,000 inpatient stays) from Northwestern Memorial Hospital (NMH) and prove the plausibility of inferred clusters of phenotypic patterns for bundled care via a survey with administrative and clinical experts. We further prove correlations of phenotypic patterns within each cluster via disease associations published in the literature.

## Research Design and Methods

The general framework is composed of four parts: i) a *workflow pattern inference module*, which is based on the electronically documented actions of EMR users, ii) a *phenotypic pattern inference module*, based on patient-specific clinical phenomena indicated in an EMR (e.g., diagnosis codes), iii) an *association module*, which infers clusters of phenotypic patterns according to their sharing workflow patterns and iv) an *evaluation module*, including surveys from administrative and clinical experts to determine if inferred clusters of phenotypic patterns could be managed in a bundle way.

We begin with a high-level overview of the models and then proceed with a deeper dive into each component. The general relationships between the workflow module, phenotypic model and association modeling algorithm are depicted in Figure 1.

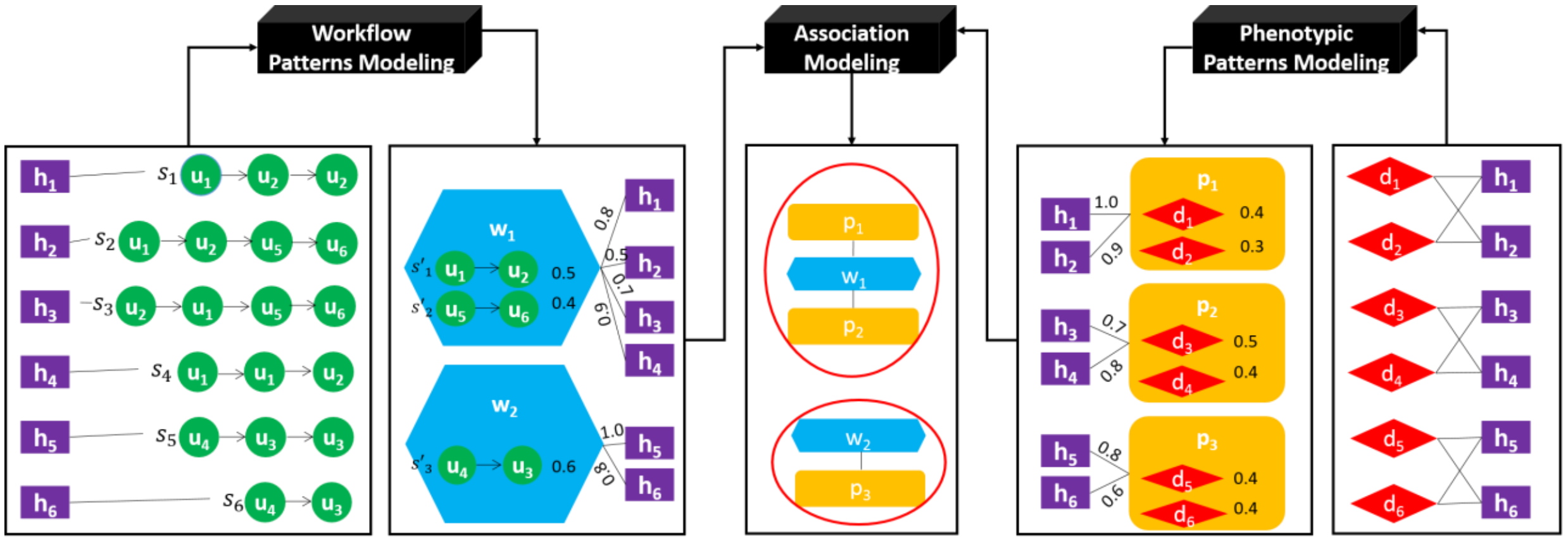

(a) Patients defined by a set of sequences

(b) Workflow patterns defined by probability distributions for a set of subsequences (left) and patients characterized by workflow patterns with likelihoods (right)

(c) Clusters of phenotypic patterns inferred via their sharing workflow patterns

(d) Phenotypic patterns defined by probability distributions for a set of diagnoses (right), and patients characterized by phenotypic patterns with likelihoods (left)

(e) Patients defined by a set of diagnoses

**Figure 1.** A high-level architecture for discovering associations between clinical workflows and phenotypes, which are further leveraged to infer clusters of phenotypes. (Legend: $u$ = *EMR user*, $h$ = *EMR patient*, $d$ = *diagnosis*, $p$ = *phenotypic pattern* and $w$ = *workflow pattern*)

Let $H = \{h_1, h_2, \cdots, h_n\}$ be the set of patients, $S = \{s_1, s_2, \cdots, s_n\}$ be the set of action sequences (issued by the users of EMRs) and $D = \{d_1, d_2, \cdots, d_l\}$ be the set of clinical phenomena (e.g., diagnosis codes). Each patient $h_\mathrm{i}$ in $H$ is defined as a sequence $s_i$ in $S$ (as shown in Figure 1a) and a collection of clinical phenomena in $D$ (as shown in Figure 1e). The set of workflow patterns $W = \{w_1, w_2, \cdots, w_\mathrm{k}\}$ (Figure 1b-left) and phenotypic patterns $P = \{p_1, p_2, \cdots, p_q\}$ (Figure 1d-right) are learned from *S* and *D*, respectively. Specifically, a workflow pattern $w_i$ is defined as a probability distribution over a set of subsequences in $S' = \{s'_1, s'_2, \cdots, s'_q\}$ (Figure 1b-left). $s'_\mathrm{i}$ is defined as a subsequence that is frequently occurring across the sequences in *S*. Similarly, a phenotypic pattern $p_j$ is a probability distribution over a set of diagnoses (e.g., see the three patterns in Figure 1d-right).

A patient is explained by their affinity to workflow and phenotypic patterns through $\varphi_W$ (Figure 1b-right) and $\varphi_P$ (Figure 1d-left), respectively. For instance, as shown in Figure 1b-right, workflow pattern $w_1$ has a probability of 0.8 of explaining the affinity between the sequence of patient $h_1$ and $w_1$. The strength of association between workflow and phenotypic patterns is summarized in a matrix $R_{|W|\times|P|}$ and is rooted in the common set of patients they explain. The collections of phenotypic patterns are inferred via the associations between the phenotypic patterns and the workflow patterns (as shown in Figure 1c).

To focus on the information learned from the EMR, in this study, we rely on existing inference algorithms to learn workflow and phenotypic patterns. To orient the reader, we briefly review the algorithms, but refer the reader to [25] and [26] for the details.

### *Workflow Pattern Inference Algorithm*

The workflow inference algorithm [25] infers workflow topics, where each topic refers to a workflow pattern, $W = \{w_1, w_2, \cdots, w_k\}$ from sequences $S'$ via a modified Latent Dirichlet Allocation (LDA) algorithm [27-28]. Briefly, the set of workflow topics $W$ is inferred from a matrix $R_{|H|\times|S'|}$. Here, $R_{|H|\times|S'|}(i,j)$ corresponds to the number of times a subsequence $s'_j$ was included within a patient sequence $s_i$. $\varphi_W$ corresponds to a matrix that specifies the likelihoods that the patients' sequences in $S$ are explained by the topics in $W$. Figure 1b-right depicts examples of the probabilities of patients' sequences being explained by workflow topics.

It is often the case that the fitness of an LDA model, and thus the number of topics $k$, is determined through an information theoretic measure, such as perplexity [27-28]. However, in our situation, we aim to determine the value that maximizes the separation between the workflow topics, which is more semantically meaningful for workflows. As such, we set $k$ by minimizing the average covariance between the workflow topics (details in [25-26]).

***Phenotypic Pattern Modeling Algorithm***

The phenotypic pattern inference algorithm [26] infers phenotypic topics $P = \{p_1, p_2, \ldots, p_q\}$ also via a modified LDA method. Briefly, the set of phenotypic topics $P$ is inferred from a matrix $R_{|H|\times|D|}$. Here, $R_{|H|\times|D|}(i,j)$ corresponds to the number of times that diagnosis code $d_j$ was assigned to patient $h_i$. Figure 1d-right depicts examples of three phenotypes as topics with two diagnoses. $\varphi_P$ corresponds to a matrix that specifies the likelihoods that patients are explained by the topics in $P$. Figure 1d-left depicts examples of the probabilities of patients' conditions being explained by phenotypic topics. We use the same strategy invoked for workflow topics to set the number of topics for phenotypic topics, which we denote as $q$ [25-26].

***Measuring Associations***

Each workflow and phenotypic topic is leveraged to explain the patients (Figure 1b and Figure 1d). We use the patients they explain in common to measure their association. Specifically, the degree of association between a workflow topic $w_i$ and a phenotypic topic $p_j$ is measured as the cosine similarity of their respective vectors:

$$\boldsymbol{Assoc}(w_i, p_j) = \frac{\varphi_W(i) \cdot \varphi_{P(j)}}{|\varphi_W(i)||\varphi_{P(j)}|}, \quad (1)$$

where $\varphi_W(i)$ is a vector specifying the distribution of probabilities that a workflow topic $w_i$ explains each of the patients. For instance, as shown in Figure 1c, the first workflow explains four patients with the following vector of probabilities ($\langle h_1, 0.8\rangle, \langle h_2, 0.9\rangle, \langle h_3, 0.7\rangle, \langle h_4, 0.6\rangle$). Similarly, $\varphi_P(j)$ is a vector specifying the distribution of probabilities that a phenotypic topic $p_j$ explains each of the patients. For instance, as shown in Figure 1d, the first phenotypic topic explains four patients with a vector of probabilities ($\langle h_1, 1.0\rangle, \langle h_2, 0.9\rangle, \langle h_3, 0\rangle, \langle h_4, 0\rangle$). According to Equation (1), the association between the first workflow and phenotypic topic $Assoc(w_1, p_1)$ is 0.7891.

Our goal is to infer clusters of phenotypic patterns that share similar workflow patterns. We suspect that each cluster would be a candidate for bundled care and management under similar workflows. Thus, we use a community detection algorithm [29] to infer clusters of phenotypic topics via their associations with workflow topics. We guided the algorithm using a heuristic that is based on the optimization of modularity [30], which is efficient (in running time) and effective (in quality of communities) for weighted and undirected graphs. Clusters with high modularity have dense connectivity of phenotypic and workflow topics within clusters and sparse connectivity across clusters.

### ***Plausibility Evaluation for Bundled Care***

We investigated if the clusters of phenotypic topics are appropriately managed in a bundled way. To do so, we designed a survey that consisted of paired ⟨*inferred*, *random*⟩ clusters of phenotypic topics, which we asked administrative and clinical experts to review for appropriateness in terms of bundled care. Each inferred phenotypic topic was represented as a list of the diagnoses (e.g., diagnostic codes) that exhibit the largest probabilities for a specific topic. A random cluster of phenotypic topics was generated by randomly selecting a number of phenotypic topics, and the number was set to be the same with the number of phenotypic topics within the inferred cluster. Each randomly selected phenotypic topic was also represented as a list of the diagnoses. Each random cluster was fixed to contain the same number of diagnoses as its inferred counterpart.

**Survey questions and analysis.** We recruited a set of experts to answer questions of the following form, "*To what extent do you believe health conditions in the displayed group can be managed in a bundled way?*" For each question, we provided five candidate answers (in the form of *Not At All Likely*, *Slightly Likely*, *Moderately Likely*, *Very Likely* and *Completely Likely*). To perform hypothesis testing, we converted these answers into values in the range 0 to 1 (e.g., *Not* = 0, *Slightly* = 0.25, *Moderately* = 0.5, *Very* = 0.75, and *Completely* = 1). Further details about the survey design, including the specific questions, are provided in online Appendix A.

Given the responses, we conducted a series of hypothesis tests, each of which can be summarized as: "*For a given pair of ⟨inferred, random⟩ clusters of health conditions, experts can distinguish the inferred from the random in terms of bundled care*". We applied a linear regression model and analysis of variance (ANOVA) [31] to test the significance of difference at the 95% confidence level.

To achieve power of 0.8 with a standard deviation of 0.4 in the difference in experts' scores for inferred and random clusters, the required sample size was five respondents. As such, we invited five knowledgeable professionals with a diverse array of expertise (e.g., HCO management, internal medicine, and emergency care). Each participant was emailed an introduction to the goals of the research and a link to access a REDCap survey [32]. The response rate was 100% because all respondents agreed to participate in the survey beforehand.

## Experimental Design

### *Dataset*

This study focused on four months of inpatient EMR data from Northwestern Memorial Hospital (NMH), which was collected in 2015. In this data, an event corresponds to an instance of a chart access, each of which is associated with the user's job title and a user-designated reason for the access. There were 1,138,317 total access events distributed over 16,569 patient processes. These events were generated by users with 144 job titles. Additionally, each patient was associated with a set of ICD-9 codes assigned after discharge from the hospital. The total number of unique ICD-9 codes for this set of patients is 4,543.

In recognition of the fact that multiple ICD-9 codes may describe the same clinical phenomena [33-34], various phenotyping investigations (e.g., [35-36]) have adopted alternative vocabularies for the secondary analysis of EMRs, such as the Phenome-Wide Association Study (PheWAS) vocabulary [15]. PheWAS codes correspond to groups of ICD-9 codes more closely match clinical or genetic understandings of diseases and reduce variability in identifying diseases. Based on this expectation, we translated a patient's ICD-9 codes to PheWAS codes, which compressed the space into 1,374 unique PheWAS codes.

### *Number of Topics*

The number of workflow and phenotypic topics were determined by minimizing the similarity over the range of 15 to 35 possible topics. The similarity was minimized for each set of topics when $k = q = 25$. At this point, the workflows and phenotypes exhibited a minimum similarity of 0.003 and 0.031, respectively.

## Results

To provide context for the findings, we begin with a depiction of the learned workflow and phenotypic topics. Next, we report on the clusters of phenotypic topics and the extent to which they were deemed plausible for bundled care by experts and had face validity according to evidence in the published literature.

### *Learned Workflow and Phenotypic Topics*

Recall that each workflow and phenotypic topic is expressed as a probability distribution over terms (i.e., subsequences of actions and PheWAS codes, respectively). To illustrate each topic succinctly, we depict the 10 terms with the largest probabilities. This cutoff was selected because the terms beyond this point had a negligible contribution to the probability mass for the affiliated topic. Specifically, these terms contributed probabilities that were smaller than 0.01.

We use ProM [37], a software tool for process mining, to visualize workflow topics as a directed graph. The graphs for all 25 workflow topics and their corresponding top 10 subsequences are provided in Appendix B. To orient the reader to workflow topics, we list workflow topic 15, consisting of two loops, as an example in Figure 2.

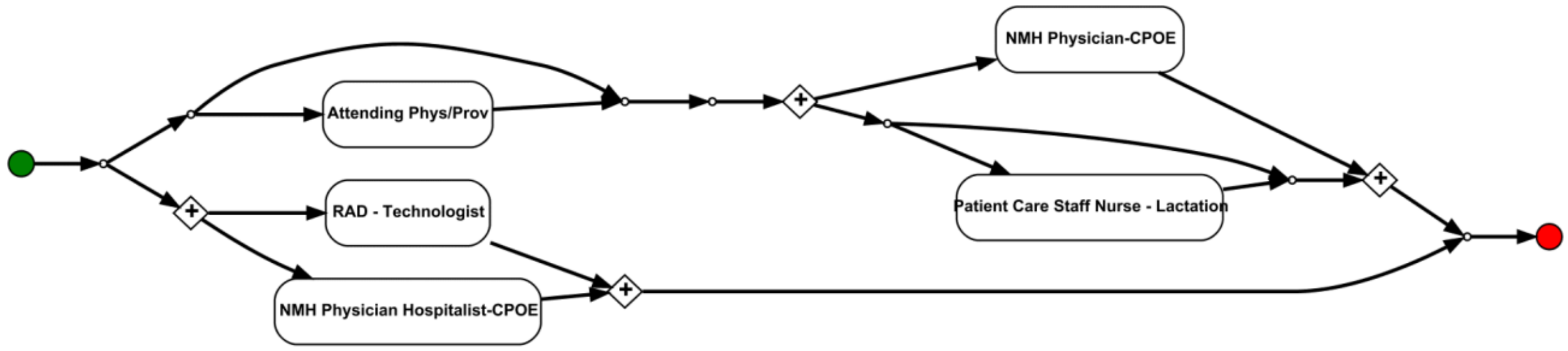

**Figure 2.** The directed graph of an echocardiography-based prenatal workflow. This visualization is based on the 10 subsequences with largest probabilities for the workflow topic. Note that, in this diagram, a pair of + symbols represents the beginning and ending of a loop.

The first loop resides between a *Radiology Technologist* (*RAD*) and an *NMH Physician Hospitalist invoking Computerized Physician Order Entry* (*CPOE*). Based on consultation with the experts, this loop was deemed to likely be associated with the process of an echocardiography, where a physician approves the quality of a radiological report or participates in the peer review process of a report. The second loop resides between an *NMH Physician CPOE* and a *Patient Care Staff Nurse-Lactation*. This loop is likely associated with a primary physician and staff nurse responsible for an inpatient's care associated with obstetrics.

Each phenotypic topic is expressed as a probability distribution over approximately 1,300 PheWAS codes. The top 10 PheWAS codes, and their associated probabilities, for each phenotypic topic is provided in Appendix C. Our author experts provided informal labels to summarize each of phenotypic topics. To better understand the phenotypic topics, we provide an example of the topics with the label of childbirth in Table 1. This topic shows that interventions are required for complicated pregnancies and delivery associated problems (e.g., short gestation, endocrine and metabolic disturbances of fetus or newborn).

**Table 1.** The top 10 PheWAS codes in a phenotypic topic that are the most indicative of childbirth.

| PheWAS Code | Description | Probability |
|---|---|---|
| 1010 | Tests associated with child birth | 0.25 |
| 637 | Short gestation; low birth weight; and fetal growth retardation | 0.18 |
| 656 | Other perinatal conditions | 0.16 |
| 656.1 | Perinatal jaundice; isoimmunization | 0.10 |
| 651 | Multiple gestation | 0.05 |
| 656.3 | Endocrine and metabolic disturbances of fetus and newborn | 0.05 |
| 747.11 | Cardiac shunt; heart septal defect | 0.05 |
| 656.2 | Other respiratory conditions of fetus and newborn | 0.02 |
| 647 | Infectious & parasitic conditions complicating pregnancy | 0.02 |
| 747.13 | Congenital anomalies of great vessels | 0.01 |

### *Clusters of Phenotypic Topics and Associated Workflow Topics*

The modularity of the clusters of phenotypic topics was 0.62 in a [0,1] range. This indicates that the phenotypic topics and workflow topics within each cluster exhibited strong associations, while they exhibited weak associations between clusters. Figure 3 depicts the four inferred clusters of phenotypic topics (shown in blue, green, purple and red) and their affiliated workflow topics.

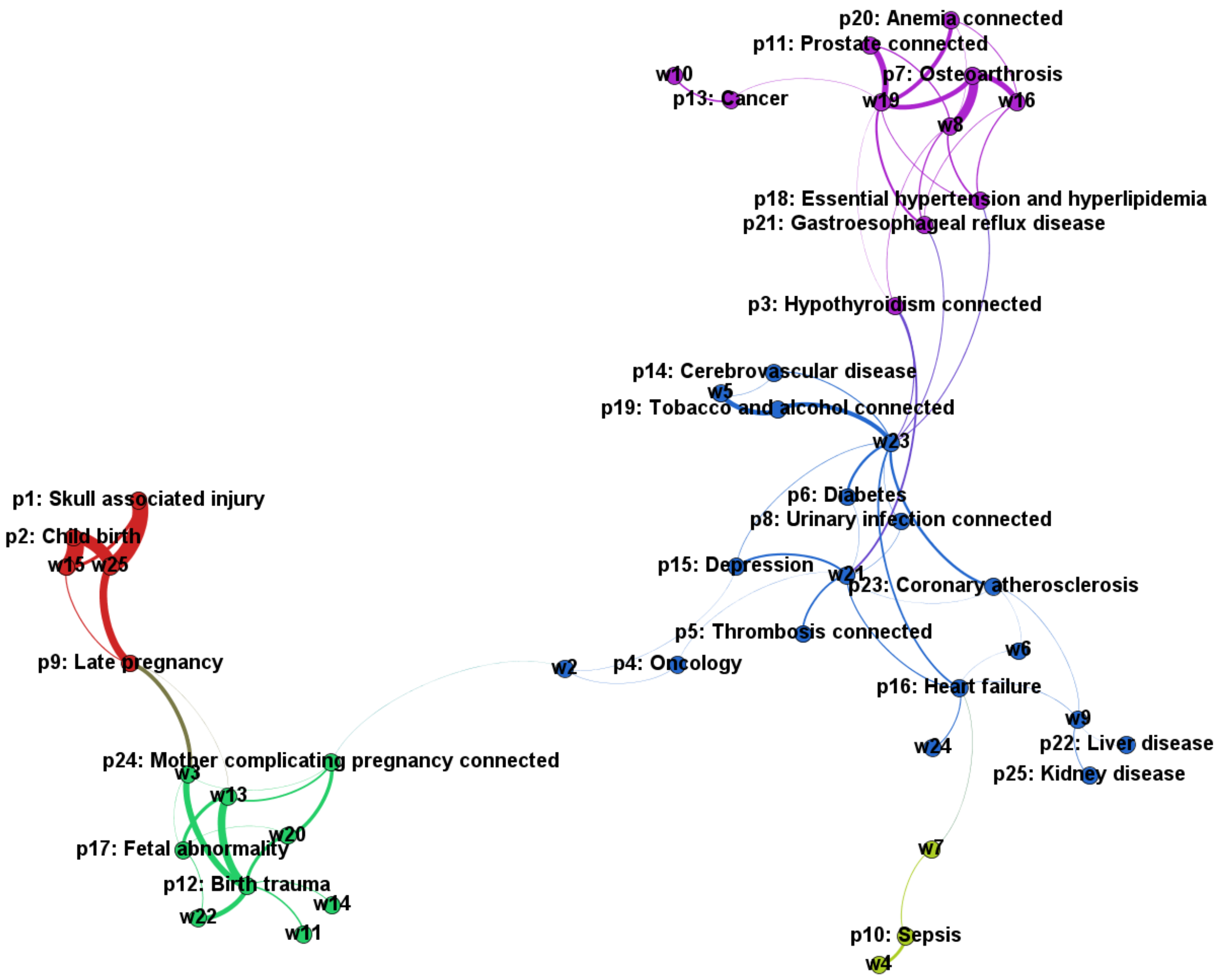

**Figure 3.** Four clusters of phenotypic topics inferred via their sharing workflow topics. The edges represent the association strength between phenotypic and workflow topic. The wider the edge, the stronger associations between phenotypic and workflow topics. (Legend: *p* = *phenotypic topic* and *w* = *workflow topic*)

Cluster $c_1$ (in green) is associated with fetal abnormality; Cluster $c_2$ (in red) is associated with late pregnancy; Cluster $c_3$ is associated with prostate problems and its corresponding complications (in purple); while cluster $c_4$ is complex, but is associated with various chronic problems, including cerebrovascular disease, coronary atherosclerosis, congestive heart failure (CHF), diabetes, and kidney failure (blue).

To gain a deeper understanding of the inferred clusters and their associated workflow patterns, let us consider $c_1$ as an example. The health conditions affiliated with $c_1$ are the following phenotypic topics:

$p_{12}$: *Birth trauma*,

$p_{17}$: *Fetal abnormality*, and

$p_{24}$: *Mother complicating pregnancy*,

which were associated with care patterns that incorporated the following workflow topics:

$w_3$: *Interactions between physicians and staff nurses,*

$w_{11}$: *Interactions between physicians, anesthesiologists, advanced practice clinicians and pharmacists,*

$w_{13}$: *Interactions between physicians and unit secretaries,*

$w_{14}$: *Interactions between physicians, anesthesiologists and staff nurses,* and

$\mathrm{w}_{22}$: *Interactions between physicians, radiologists and unit secretaries*.

This suggests that pregnancy complications (e.g., fetal abnormality and mother complicating pregnancy) are managed in a bundled way, requiring communication between various clinicians, obstetricians, anesthesiologists, radiologists, nurses, pharmacists, and administrative assistants.

***Plausibility of Phenotypic Clusters for Bundled Care***

The results of the plausibility survey are provided in Table 2. It can be seen that the experts always scored the inferred clusters as the more plausible for bundled care. All four clusters were statistically significantly higher than the randomized cluster in terms of the respondents' scores (using a 95% confidence interval). This suggests that the phenotypic clusters associated with fetal abnormality, late pregnancy, prostate problems and CHF are plausible candidates for bundled care.

Additionally, to orient the reader to each phenotypic cluster, we provide each of them, along with an informal summary from our author experts, in Table 2.

**Table 2.** Survey results for the knowledgeable experts (*n* = 5) regarding the plausibility of the inference that phenotypic patterns in each cluster can be managed in a bundled manner. Each cluster of phenotypic patterns are represented by a list of PheWAS codes and a brief summary. Each row reports the distance between the Likert score of the inferred phenotypic cluster and its randomized counterpart. Note that a positive distance indicates the inferred cluster received a higher Likert score. (* = statistical significance at the 0.05 confidence level)

| Cluster | PheWAS Codes and Descriptions | Likert Score Difference | P-value |
|---|---|---|---|
| **Informal Description:** Fetal abnormality could lead to complicating pregnancy and additional delivery problems (e.g., fetal distress), which requires interventions such as birth trauma service. | | | |
| $C_1$ | 649 Other conditions of the mother complicating pregnancy<br>652 Malposition and malpresentation of fetus or obstruction<br>654 Abnormality pelvic soft tissues & organs complicating pregnancy<br>658 Problems associated with amniotic cavity and membranes<br>659 Indications for care or intervention related to labor and delivery NEC<br>663 Umbilical cord complications during labor and delivery<br>665 Obstetrical/birth trauma | 0.95 | $6.09 \times 10^{-8*}$ |
| **Informal Description:** Late pregnancy might suggest a larger size infant requiring intervention (e.g. use of suction or forceps) which may cause temporary skull injuries. | | | |
| $C_2$ | 637 Short gestation; low birth weight; and fetal growth retardation<br>645 Late pregnancy and failed induction<br>649 Other conditions of the mother complicating pregnancy<br>656 Other perinatal conditions<br>656.1 Perinatal jaundice/isoimmunization<br>665 Obstetrical/birth trauma<br>819 Skull fracture and other intracranial injury<br>1010 Other tests<br>1008 Internal injury to organs | 0.95 | $6.09 \times 10^{-8*}$ |
| **Informal Description:** Anemia and hypogonadism are often considered complications of prostate cancer and can lead to bone loss. When the thyroid does not produce a sufficient amount of hormones, it can cause lower esophageal sphincter dysfunction. This allows stomach contents and digestive juices to enter the esophagus, which may lead to gastroesophageal reflux disease. | | | |
| $C_3$ | 244 Hypothyroidism<br>272.1 Hyperlipidemia<br>276.14 Hypopotassemia<br>285.9 Anemia NOS<br>327.32 Obstructive sleep apnea<br>401.1 Essential hypertension<br>495 Asthma<br>530.11 Gastroesophageal Reflux Disease<br>600 Hyperplasia of prostate<br>740.1 Osteoarthritis; localized | 0.65 | $2.80 \times 10^{-4*}$ |
| **Informal Description:** Cerebrovascular disease and coronary atherosclerosis are the most common cause of congestive heart failure (CHF); smoking and diabetes are associated with all of the three diseases. Depression is associated with coronary disease. The liver test abnormality and some renal failure may be seen in CHF. | | | |
| $C_4$ | 250.2 Type 2 diabetes<br>272.1 Hyperlipidemia<br>286.5 Hemorrhagic disorder due to intrinsic circulating anticoagulants<br>296.2 Depression<br>316 Substance addiction and disorders<br>318 Tobacco use disorder<br>401.1 Essential hypertension<br>401.22 Hypertensive chronic kidney disease<br>427.21 Atrial fibrillation<br>428 Heart failure<br>428.1 Systolic/diastolic heart failure<br>433.31 Transient cerebral ischemia<br>452 Venous embolism & thrombosis<br>585.3 Chronic renal failure<br>591 Urinary tract infection<br>707.1 Decubitus ulcer | 0.70 | $7.04 \times 10^{-5*}$ |

*Face Validity of Phenotypic Clusters according to Evidence in the Published Literature*

While the phenotypic clusters were deemed plausible for bundled care from care management perspective, we did not investigate if the health conditions within such clusters were clinically related. If appropriateness could be confirmed from both a care process and a clinical perspective, we anticipate that the identified clusters of phenotypic patterns would be better received by HCO administrators.

Towards this goal, we performed an investigation into evidence for the inferred clusters of phenotypic patterns. Notably, we found evidence for each cluster. A summary of the evidence is shown in Table 3. For instance, within cluster $c_3$, bone loss is known to be caused by hypogonadism following prostate cancer [38]. Furthermore, acid reflux is known to be affiliated with thyroid problems [39].

**Table 3.** Evidence from the literature supporting the face validity of phenotypic patterns within each inferred cluster.

| Cluster | Evidence of Associations in the Literature |
|---|---|
| $c_1$ | • Birth trauma associated with fetal big size and fetal distress [40]<br>• Trauma in pregnancy [41-42] |
| $c_2$ | • Late pregnancy and child birth [43]<br>• Mode of delivery in nulliparous women has an effect on neonatal intracranial injuries [44]<br>• Most fetal injuries occur in late pregnancy [45] |
| $c_3$ | • Bone loss following hypogonadism with prostate cancer [38]<br>• The acid reflux-thyroid connection [39]<br>• Anemia associated with advanced prostate cancer [46] |
| $c_4$ | • Tobacco and alcohol usage had increased risk of mortality for cerebrovascular disease and liver disease [47]<br>• Thrombotic complications in heart failure [48-49]<br>• Associations among diabetes, kidney disease, and cardiovascular disease [50] |

## Discussion

### *Main Findings*

This pilot study has several notable implications. First, the findings suggest that HCOs have an opportunity to leverage inferred phenotypic patterns, along with their affiliated workflow patterns, to identify (or refine) bundled care models. For instance, for patients near childbirth, their conditions may be affiliated with phenotypic topics: $p_{12}$: *Birth trauma*, $p_{17}$: *Fetal abnormality*, and $p_{24}$: *Mother complicating pregnancy*, which were associated with care patterns incorporating workflow topics: $w_3$: *associated with physicians and care staff nurses*, $w_{14}$: *associated with anesthesiologists*, and $w_{22}$: *associated with radiologists*. Second, the associations between workflow and phenotypic topics provide an opportunity for HCOs to manage patients and conduct resource allocation more efficiently. For instance, if the volume of patients associated with childbirth increases, HCOs could dedicate a larger amount of resources to workflow topics $w_3$, $w_{14}$ and $w_{22}$.

### *Limitations and Next Steps*

Despite the merits of our findings, there are several limitations that we wish to highlight for future investigations. First, this study focused on the development of a methodology to infer general collections of phenotypic patterns that share similar workflow patterns according to EMR system utilization. However, we did not validate the clinical meanings (e.g., semantic contexts) for each of the inferred phenotypes nor their workflow patterns. If such phenotypic and workflow patterns are to be used in care management applications, their semantic meanings will require further interpretation by administrative experts.

Second, while all four phenotypic clusters were deemed plausible for bundled care, several associations within congestive heart failure cluster $c_4$ were not clear to the experts. Specifically, there are a number of reasons why renal failure and liver diseases might co-occur in a patient, such that this cluster may be too general in nature. In this respect, our study indicates health conditions have the potential to be managed in a bundled way, but what precisely should be managed is an open question and will require guidance by process management experts.

Third, we acknowledge that this is a pilot only, which focuses on a case study of four months of data from one HCO.  As such, we only found four clusters of phenotypic patterns, which were showed to be suggestive for bundled care. It is unknown if the proposed strategy is directly generalizable to other healthcare systems to find more clusters of health conditions, which could have high opportunities to be managed in a bundled manner.

## Conclusions

In this paper, we introduced a data-driven framework to mine EMRs for clusters of health conditions that might benefit from bundled care. We evaluated the approach with four months of inpatient data from a large hospital system and found four clusters of phenotypic patterns, which were deemed plausible for bundled care by knowledgeable experts and evidence in the literature. We anticipate working with process management and clinical experts to assess the workflow patterns affiliated with each inferred cluster to figure out how these patterns can be incorporated together to provide bundled care. Furthermore, we plan to test the performance and efficacy of such the framework in other healthcare systems with more data.

## Appendices

Appendix A: Survey questions.

Appendix B: Workflow topics, each of which is represented by its top 10 subsequences and visualized as a process graph via Business Process Model and Notation (BPMN) in ProM.

Appendix C: Phenotypic topics, each of which is represented by its top 10 PheWAS codes.

**Funding**

This research was supported, in part, by the National Institutes of Health under grants R00LM011933 and R01LM010685.

**Competing Interests Statement**

The authors have no competing interests to declare.

**Contributors**

YC performed the data collection and analysis, methods design, hypotheses design, experiments design, evaluation and interpretation of the experiments, and writing of the manuscript. AK and DL performed data collection, evaluation and interpretation of the experiments and writing of the manuscript. CI, SO, and JB performed evaluations of inferred clusters of phenotypes, and writing of the manuscript. BM performed the data collection and analysis, evaluation and interpretation of the experiments, and writing of the manuscript.

**Acknowledgements**

The authors thank Daniel Schneider and Prasanth Nannapaneni for gathering and supplying the de-identified data from Northwestern Memorial Hospital analyzed in this investigation.